\documentclass[aps,pra,reprint,superscriptaddress,floatfix,nofootinbib,amsmath,amssymb,longbibliography]{revtex4-2}

\usepackage{graphicx}
\usepackage{mathdots}
\usepackage{epstopdf}
\usepackage{hyperref}
\usepackage{tikz}
\usepackage{physics}
\usepackage{multirow}
\usepackage{amsmath,amssymb,amsfonts}
\usepackage{amsthm}
\usepackage{mathrsfs}
\usepackage[title]{appendix}
\usepackage{xcolor}
\usepackage{textcomp}
\usepackage{booktabs}
\usepackage{algorithm}
\usepackage{algorithmicx}
\usepackage{algpseudocode}
\usepackage{listings}
\usepackage{float}
\usepackage{needspace}
\usepackage{soul}

\begin{document}

\title{Practical scheme for efficient distillation of GHZ states}

\author{\'Aron Rozgonyi}
\email{aron.rozgonyi@pqi.pt}
\affiliation{PQI -- Portuguese Quantum Institute, Portugal}
\affiliation{Physics of Information and Quantum Technologies Group, Centro de F\'{\i}sica e Engenharia de Materiais Avan\c{c}ados (CeFEMA), Portugal}
\affiliation{ELTE E\"otv\"os Lor\'and University, H-1117 Budapest, Hungary}
\affiliation{HUN-REN Wigner Research Centre for Physics, H-1525 Budapest, Hungary}

\author{G\'abor Sz\'echenyi}
\affiliation{ELTE E\"otv\"os Lor\'and University, H-1117 Budapest, Hungary}
\affiliation{HUN-REN Wigner Research Centre for Physics, H-1525 Budapest, Hungary}

\author{Orsolya K\'alm\'an}
\affiliation{HUN-REN Wigner Research Centre for Physics, H-1525 Budapest, Hungary}

\author{Tam\'as Kiss}
\affiliation{HUN-REN Wigner Research Centre for Physics, H-1525 Budapest, Hungary}


\begin{abstract}
Generating shared Greenberger-Horne-Zeilinger (GHZ) states remains a challenge due to noise during the distribution process. The imperfections caused by noise can be mitigated through entanglement distillation, a critical component of quantum communication protocols.
We present an efficient local operation and classical communication (LOCC) scheme for distilling GHZ states of tripartite systems affected by both coherent and incoherent errors. Our method utilizes an iterative process that applies a postselection-based nonlinear LOCC transformation on input tripartite states pairwise to enhance the entanglement of one of them. 
We present a double-iteration protocol by establishing a mathematical framework for the transformation processes, with particular emphasis on finding  specific unitary operations which can correct small arbitrary errors in the initial states. Our protocol can probabilistically transform 4 input copies of noisy three-qubit GHZ states into a single, almost-perfect GHZ state, eliminating noise to first order, in analogy with the scheme proposed in~\cite{kalman2025unambiguous} for bipartite Bell-state distillation. Moreover, we  analytically derive a corresponding no-go theorem by demonstrating that a similar single-iteration protocol acting on only 2 inputs is impossible.
Notably, our protocol corrects small arbitrary distortions in GHZ states, converging superexponentially to a pure GHZ state while maintaining operational simplicity, thereby supporting its feasibility for practical quantum computing applications. The possibility of eliminating errors to first order, together with the fact that only four inputs suffice, makes our protocol especially beneficial and practical for applications in quantum communication schemes, distributed quantum systems, and in general, for the future quantum internet.  
\end{abstract}

\maketitle

\section{Introduction}

Quantum entanglement provides an advantage to quantum systems over classical ones for various applications in the fields of communication, computation, and sensing. Multipartite entangled states are essential in many areas of quantum information processing~\cite{multipartite_review}. This type of entanglement is fundamentally different from bipartite entanglement, its quantification and characterization is not a straightforward generalization of the bipartite case~\cite{Szalay}. Moreover, it is a potential asset in several fields, such as distributed and measurement based quantum computing, quantum error correction,  
and quantum internet applications. It is also highly relevant in quantum sensing and metrology~\cite{toth2012multipartite}. 

The Greenberger-Horne-Zeilinger (GHZ) state~\cite{greenberger1989going} is a maximally entangled tripartite or multipartite state. 
GHZ states have been realized in different platforms, including photons~\cite{pan2001experimental,PhysRevLett.132.130604,pont2024high}, trapped ions~\cite{roos2004control}, semiconducting qubits~\cite{nogueira2021dynamic}, and superconducting qubits~\cite{dicarlo2010preparation}. The preparation of high-fidelity GHZ states on distributed systems remains a challenge, albeit being essential for quantum networking, scaling up architectures, or distributed quantum computing. In a recent realization, three trapped ions in separate traps have been prepared in a GHZ state using a photonic interconnect in a direct manner, i.e., by three-photon Hong-Ou-Mandel interference \cite{goetting2026tripartiteentanglement}.
Non-perfect generation or noisy transmission results in lower fidelity GHZ states that are insufficiently entangled for applications. Therefore it may be necessary to purify or distill these states~\cite{10908198}.

Entanglement distillation is an operation to convert a number of non-maximally  entangled input states into a smaller number of states with higher entanglement, using only local operations and classical communication (LOCC). Distillation of multipartite entangled states, especially, GHZ states has attracted considerable attention~\cite{PhysRevA.57.R4075,GHZ-purification-paritycheck,W-distillation,huang2014distillation,loccnet,Krastanov_2,qin2023efficientmultipartiteentanglementpurification,Dür_2007,dur-ghz-network,PhysRevLett.91.107903,PhysRevA.71.012319,Kruszynska2006all,Kruszynska2006cost,Wallnofer2016repeaters,MiguelRamiro2023W,Stloukalova2026LEP}. 
For current practical applications it is highly desirable to minimize the number of processed input states as well as the number local operations~\cite{fang2019non,Krastanov_2,Krastanov_2025}. From a practical point of view it may also be advantageous to apply the same local operations and measurements in an iterative manner. 

In a recent proposal for two-qubit distillation a scheme was presented with which Bell states can be prepared universally and unambiguously from a small number of arbitrary input states containing moderate noise~\cite{kalman2025unambiguous}.
The behavior of the protocol for pure input states (entanglement concentration) has since been shown to be optimal~\cite{Orthey_2026}. Another important feature of this distillation protocol is that it suppresses initial noise quadratically already after a small number of iterations (two or three depending on the type of input state). A more recent analysis points out the importance of quadratic noise suppression for a resource efficient scaling up of quantum computers~\cite{Baumer_arxiv_2026}.  
A simple generalization of the above-mentioned two-qubit scheme for three qubits is not straightforward. In previous work, we trained variational quantum circuits to develop an effective protocol for distilling GHZ states affected by white noise ~\cite{rozgonyi2024training}. The numerical results showed that, typically, even-odd oscillations occur, thus optimization may be achieved by considering every second iterative step.

In this paper we analytically investigate the possibility of generalizing the protocol in~Ref.~\cite{kalman2025unambiguous} to GHZ states. First, we prove  that this type of protocol cannot monotonically increase the fidelity by every iteration. Then, we construct double-iteration protocols by establishing a mathematical framework for the transformation processes, with particular emphasis on finding  specific unitary operations which can correct small arbitrary errors in the initial states. We show that these protocols can probabilistically transform 4 input copies of noisy GHZ states into a single, almost-perfect GHZ state, eliminating noise to first order, very similarly to the scheme of Ref.~\cite{kalman2025unambiguous}. These protocols correct both coherent and incoherent errors of the input states. Our protocols have further advantages: (i) converge superexponentially to the pure GHZ state as a function of the number of iterations, and (ii) the unitary operations have a simple gate decomposition.

The rest of the paper is arranged as follows. In Sec.~\ref{sec_setup},
we briefly review the distillation protocol of the GHZ states and formulate the conditions for achieving fast convergence. One iteration of the procedure is applied and investigated in Sec.~\ref{ref_single}. A double iteration of the protocol with different (the same) operations is studied in Sec.~\ref{ref_double_alternating} (Sec.~\ref{ref_double_uniform}). In Sec.~\ref{sec:practical}, we compare the resource cost of our protocol with that of Ref.~\cite{PhysRevA.57.R4075}. The paper is enclosed with concluding remarks in Sec.~\ref{conclusion}.

\section{Setup} \label{sec_setup}
The goal of this work is to design a distillation protocol that efficiently corrects both coherent and incoherent errors. For this purpose, we consider an ensemble of three-qubit systems, each prepared in the same quantum state. We assume that
the three qubits are shared between three different parties located at possibly distant places. In this paper, we adopt a framework of LOCC (see Fig.~\ref{fig:setup}) consisting of the following steps:

\begin{figure}[h!]
\centering
\includegraphics[width=0.9\columnwidth]{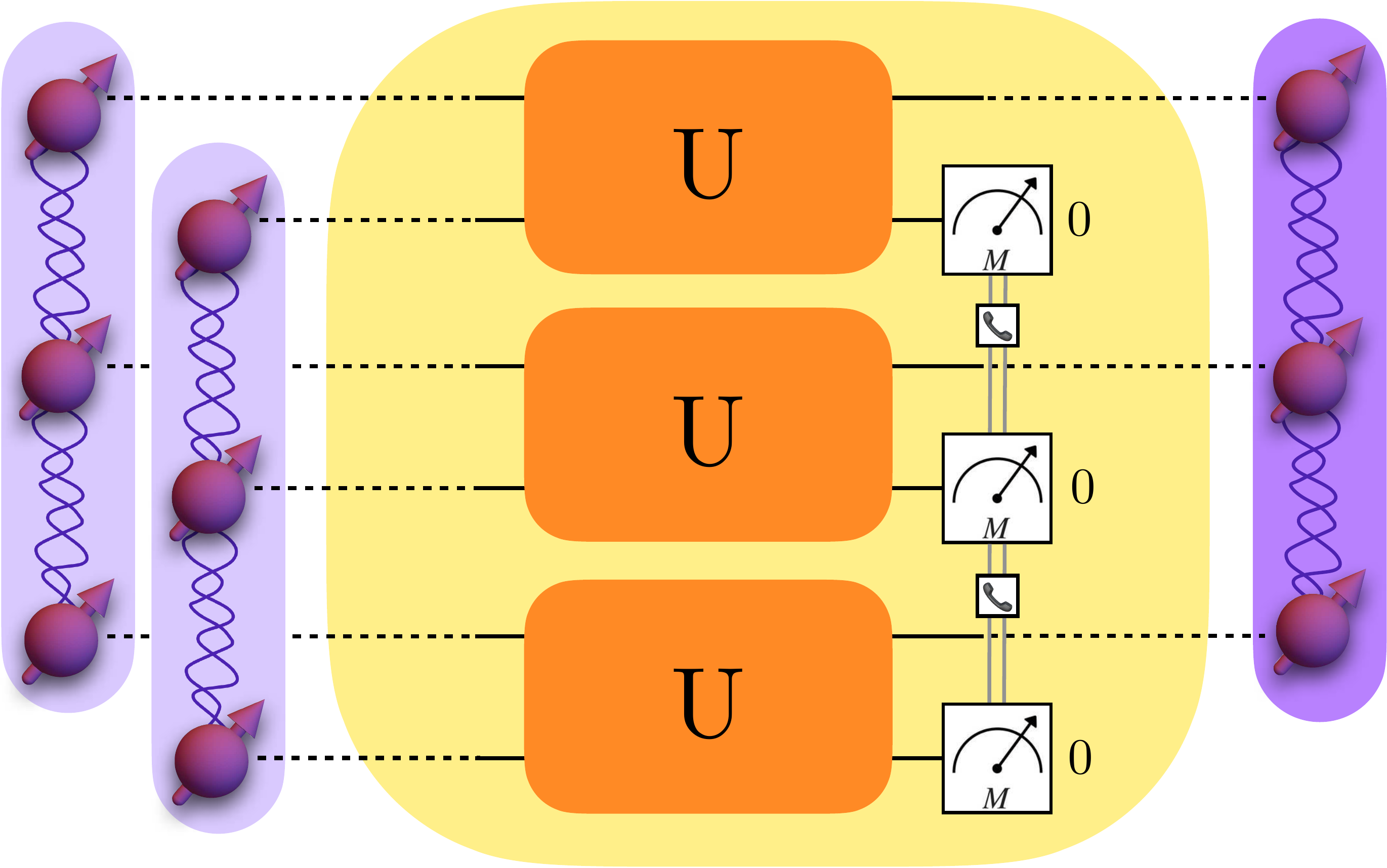}
\caption{\label{fig:setup} Schematic representation of the protocol for 3-qubit GHZ state distillation. Three distant parties apply a local unitary operator ($U$) and a subsequent measurement ($M$). They only keep the unmeasured qubits for the next iteration if all measurement outcomes are zero, which they can communicate classically.}
\end{figure}

\begin{enumerate}
    \item \textit{Distribution of qubits:}
Two copies from the ensemble are shared among three parties such that each participant receives a single qubit from each state. The qubits of one of the copies will serve as flag qubits.
    \item \textit{Local operation:} Each party performs the same unitary operation ($U$) on their two qubits locally.
    \item \textit{Measurement:} Each party performs a von-Neumann measurement on their flag qubits in the computational basis.
    \item \textit{Classical communication and post-selection:} The parties share their measurement results via classical communication channels. They keep the unmeasured qubits only if all three measurements resulted in zero. For any other measurement outcome, they discard the unmeasured qubits and repeat steps 1--3 on a new pair of states from the ensemble.
    \item \textit{Iteration of the procedure:}
The unmeasured, successfully post-selected qubits are used as inputs for the next application of the scheme with steps 1--4. 
\end{enumerate}

A single iteration of the protocol transforms the initial density matrix $\rho_\textrm{in}$ into $\rho_\textrm{out}$ by a nonlinear transformation, which depends on the unitary operator $U$. Let us denote this as
\begin{equation}
\mathcal{P}_U\left[\rho_\textrm{in}\right]=\rho_\textrm{out}.
\end{equation}
We note that if $\rho_{\mathrm in}$ is a pure state, then the scheme, due to the strict post-selection criterion that we apply (namely, only a single measurement result is accepted), guarantees that $\rho_{\mathrm out}$ will also remain a pure state. This property is unlike most distillation schemes. Our aim is to distill tripartite GHZ states
\begin{equation}
|GHZ\rangle=\frac{1}{\sqrt{2}}\left(|000\rangle + | 111\rangle \right).
\end{equation}
For the input we assume an ensemble in a slightly distorted GHZ state,
represented by the following density matrix:
\begin{equation} \label{error}
\rho_\textrm{in}=\rho_\textrm{GHZ}+\epsilon M_\epsilon,
\end{equation}
where $\rho_\textrm{GHZ}=|GHZ\rangle \langle GHZ|$ is the density matrix of the pure GHZ state, $\epsilon\ll 1$ is a small dimensionless parameter characterizing the strength of the noise, and $M_\epsilon$ is a traceless, Hermitian matrix. This framework enables us to consider both coherent and incoherent errors. 

Let us now consider these two types of errors separately.

(i) In the case of a coherent error, the initial state is a pure state, which is slightly rotated away from the perfect GHZ state:
$\ket{\Psi_{\mathrm{in}}}=\mathcal{N}\left(|GHZ\rangle+\epsilon |\Psi_\epsilon\rangle\right)$,
$\mathcal{N}$ being a normalization factor, and $\ket{\Psi_\epsilon}$ is orthogonal to the GHZ state.
If we neglect second-order terms in $\epsilon$, then the density matrix has the form given by Eq.~(\ref{error}), where
$M_\epsilon$ can be written as
\begin{equation} \label{noise_pure}
M_\epsilon=|\Psi_\epsilon\rangle\langle GHZ| + \textrm{h.c.}
\end{equation}

(ii) In the case of small incoherent noise, the initial state is a GHZ state slightly mixed with noise, which can be written in the form
$\rho_\textrm{in} = (1-\epsilon)\rho_\textrm{GHZ}+\epsilon \rho_\textrm{noise}$, where $\rho_\textrm{noise}$ is the density matrix characterizing the noise.
For such an incoherent error, $M_\epsilon$ has the form $M_\epsilon=\rho_\textrm{noise}-\rho_\textrm{GHZ}$. One widely used example is the white noise approximation, where $\rho_\textrm{noise}=\frac{1}{8}I$.

The goal of our work is to construct a protocol that suppresses both coherent and incoherent errors efficiently.
For this purpose, in what follows, we impose the condition that the protocol eliminates the considered error in linear order after a few iterations, so that the outgoing state will contain the error at most in second order: $\rho_\textrm{out}=\rho_\textrm{GHZ} + \mathcal{O}(\epsilon^2)$. Repeating such a protocol then guarantees that the outgoing state will quickly converge to the GHZ state.

\section{Single iteration protocol} \label{ref_single}

The first relevant question about this setup is whether one can construct a protocol that eliminates the above-mentioned type of errors in linear order after only a single iteration of the scheme or, in other words, whether there exists an operator $U$, which fulfills the condition
\begin{equation} \label{eq:condition1}
\mathcal{P}_U\left[\rho_\textrm{GHZ}+\epsilon M_\epsilon\right]=\rho_\textrm{GHZ} +\mathcal{O}(\epsilon^2).
\end{equation}

In what follows, we show that there is no unitary operator that would satisfy the condition written in Eq.~(\ref{eq:condition1}) for arbitrary type of noise. Nevertheless, we find some unitary operators by which a partial distillation can be achieved.

\subsection{No unitary for single-iteration distillation}

For the analytical investigation, we define the matrix elements $U_{ij}$ of the unitary operator $U$ using the basis set $|00\rangle$, $|01\rangle$, $|10\rangle$ and $|11\rangle$, where the first (second) qubit is the unmeasured (flag) qubit. Elements of any unitary matrix fulfill the conditions
\begin{equation}\label{eq:condition_unitary}
\sum_j U_{ij}U_{kj}^*=\delta_{ik},
\end{equation}
where $\delta$ is the Kronecker-delta symbol.

The statement that the GHZ state is a fixed point of the non-linear transformation, i.e.\ $\mathcal{P}_U\left[\rho_\textrm{GHZ}\right]=\rho_\textrm{GHZ}$, gives the following condition for the matrix elements
\begin{eqnarray}\label{eq:condition_fix}
\sum_{i=1}^4(U_{1i})^3&=&\sum_{i=1}^4(U_{3i})^3\neq 0,\nonumber \\
\sum_{i=1}^4(U_{1i})^2U_{3i}&=&\sum_{i=1}^4(U_{3i})^2U_{1i}=0.
\end{eqnarray}

Next, we take coherently distorted GHZ states, where $M_\epsilon$ is given by Eq.~(\ref{noise_pure}). Solving
Eq.~(\ref{eq:condition1}) for this special noise leads to the following conditions,

\begin{eqnarray}\label{eq:condition_pure}
&&U_{11}U_{31}(U_{i2}+U_{i3}) + U_{i4}(U_{12}U_{32}+U_{13}U_{33})= 0,\nonumber \\
&&U_{14}U_{34}(U_{i2}+U_{i3}) + U_{i1}(U_{12}U_{32}+U_{13}U_{33})= 0,\nonumber\\
&&(U_{11})^x\left[(U_{32})^{3-x}+(U_{33})^{3-x} \right]+\nonumber\\ &&\;+(U_{34})^{3-x}\left[(U_{12})^x+(U_{13})^x \right]=0,\nonumber \\
&&(U_{31})^x\left[(U_{12})^{3-x}+(U_{13})^{3-x} \right]+\nonumber\\ &&\;+(U_{14})^{3-x}\left[(U_{32})^x+(U_{33})^x \right]=0,\nonumber \\
&&U_{11}^x\left[(U_{12})^{3-x}+(U_{13})^{3-x} \right] +U_{14}^{3-x}\left[(U_{12})^{x}+(U_{13})^{x}\right]=\nonumber\\ &&\;= U_{31}^x\left[(U_{32})^{3-x}+(U_{33})^{3-x} \right] +U_{34}^{3-x}\left[(U_{32})^{x}+(U_{33})^{x}\right] ,\nonumber\\
&& U_{11}^3-U_{31}^{3}+ U_{34}^3-U_{14}^{3}=0,\nonumber \\
&&U_{11}^xU_{31}^{3-x}- U_{14}^xU_{34}^{3-x} = 0
\end{eqnarray}
for every $x\in\left\{1,2\right\}$ and $i\in\left\{1,3\right\}$. In the protocol, we keep the state only if the measurement outcome of the flag qubit is zero, therefore, only the matrix elements from the first and third rows appear in the conditions Eqs.~(\ref{eq:condition_fix}) and (\ref{eq:condition_pure}).

The conditions listed so far in Eqs.~(\ref{eq:condition_unitary})--(\ref{eq:condition_pure}) must be solved simultaneously. There are two kinds of solutions, in one case $|U_{12}|= |U_{33}|=1$ and all the other elements in the first and third rows are zeroes, furthermore, the relative phase of $U_{12}$ and $U_{33}$ is 0, $2\pi/3$ or $-2\pi/3$. In the other case, $|U_{13}|= |U_{32}|=1$, while all the other elements in the first and third rows are zeroes, and similarly, the relative phase of $U_{13}$ and $U_{32}$ is 0, $2\pi/3$ or $-2\pi/3$. One of these solutions is investigated in detail in the next subsection.

We have successfully derived unitary matrices for a protocol which efficiently distills pure states slightly distorted from $|GHZ\rangle$. However, if we test any of these protocols for a white-noise-affected GHZ state, then the error is not eliminated in the first order, and condition Eq.~(\ref{eq:condition1}) is violated. It means that no protocol can successfully eliminate in leading order both the coherent and incoherent errors during a single iteration step.

However, we find some unitary operators that can be used to realize partial distillation. Two of them are analyzed in detail in the upcoming two subsections.

\subsection{The CNOT-X operator}

\begin{figure}
\centering
\includegraphics[width=0.95\columnwidth]{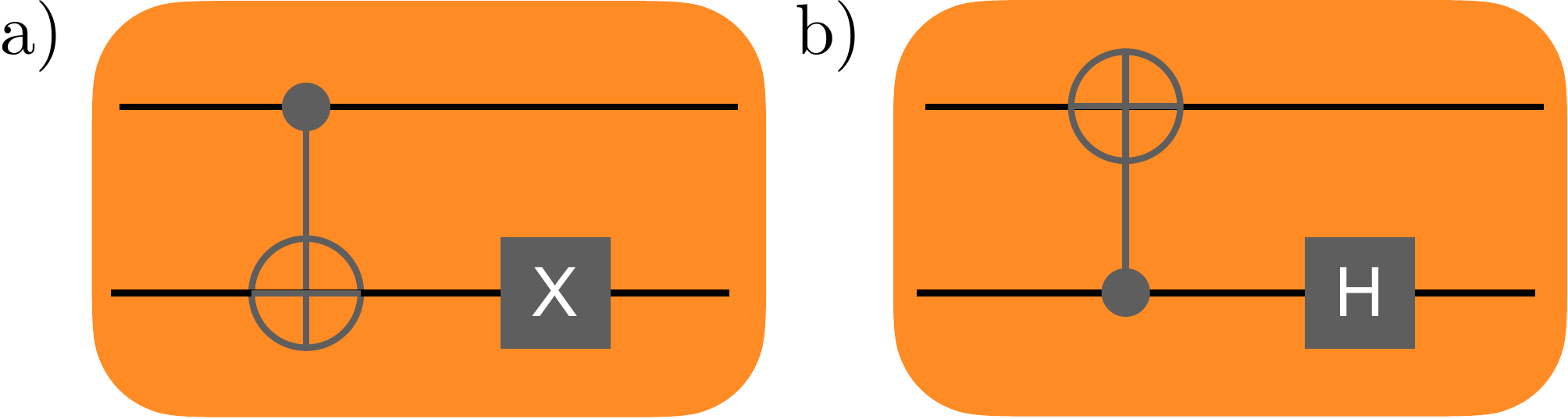}
\caption{\label{fig:decomposition} The gate decomposition of the unitary operators $U_1$ and $U_2$ used in the distillation protocols. The bottom line represents the flag qubit.}
\end{figure}

One of the unitary operators, which fulfills the conditions Eqs.~(\ref{eq:condition_unitary})--(\ref{eq:condition_pure}) reads
\begin{align}
U_1=
\begin{pmatrix}
0 & 1 & 0 & 0\\
1 & 0 & 0 & 0\\
0 & 0 & 1 & 0\\
0 & 0 & 0 & 1
\end{pmatrix}
\end{align}
in the
$\left\lbrace|00\rangle, |01\rangle,|10\rangle, |11\rangle\right\rbrace$ basis. This is called the CNOT-X operator as it can be decomposed into a CNOT gate and an X gate, as shown in Fig.~\ref{fig:decomposition}(a).

The GHZ state is a fixed point of the protocol using $U_1$. Moreover, it transforms any pure initial state of the form $|\Psi_0\rangle = \alpha |000\rangle + \beta |111\rangle$ ($\alpha,\beta \in \mathbb{C}$) exactly into the GHZ state,
\begin{equation}\label{eq:U1GHZ}
\mathcal{P}_{U_1}\left[|\Psi_0\rangle\langle \Psi_0| \right]=\rho_\textrm{GHZ}.
\end{equation}
This means that the protocol utilizing $U_1$ distills the GHZ state from any superposition state of $|000\rangle$ and $|111\rangle$ in one iteration. The success rate of the protocol, i.e., the probability that all parties measure zero on their flag qubits, reads
\begin{equation}
P(0,0,0)=2\left|\langle 000|\Psi_0\rangle\right|^2 \left|\langle 111|\Psi_0\rangle\right|^2.
\end{equation}
We note that for an ideal GHZ state as input (for which the output is again the GHZ state), the success rate is 0.5.

For a noisy GHZ state, the non-linear transformation proceeds as
\begin{equation} \label{eq:PU1}
\mathcal{P}_{U_1}\left[\rho_\textrm{GHZ}+\epsilon M_\epsilon\right]=\rho_\textrm{GHZ} +\epsilon \tilde{M}_\epsilon + \mathcal{O}(\epsilon^2),
\end{equation}
where the leading-order error reads
\begin{equation} \label{eq:Mepsilon}
\tilde{M}_\epsilon = -2 \langle GHZ^- |M_\epsilon |GHZ^-\rangle | 000 \rangle \langle 111 | + \textrm{h.c.},
\end{equation}
and the notation $|GHZ^-\rangle = \frac{1}{\sqrt{2}} \left(|000\rangle - |111\rangle\right)$ was introduced.

For pure state inputs that are slightly deviated from the ideal GHZ state, the first-order error term in the output is zero, i.e., $\tilde{M}_\epsilon=0$. This is readily verified by substituting Eq.~(\ref{noise_pure}) into Eq.~(\ref{eq:Mepsilon}) and utilizing the fact that the GHZ and the GHZ$^-$ states are orthogonal to each other.

For input states where $\langle GHZ^- |M_\epsilon |GHZ^-\rangle \neq0$, we find that the GHZ state is still a fixed point of the non-linear transformation but an unstable one. Therefore, the protocol using $U_1$ cannot distill the GHZ state from such input states in practical situations.

It is important to note that
if the input states are arbitrary but slightly distorted GHZ states, then the leading order noise term of the output $\tilde{M}_\epsilon$ contains matrix elements only in the subspace spanned by the $|000\rangle$ and the $|111\rangle$ states. This statement remains valid even if the two incoming states are not perfect replicas of each other.

\subsection{The CNOT-H operator}

The other unitary operator that we analyze in detail reads
\begin{align}
U_2=\frac{1}{\sqrt{2}}
\begin{pmatrix}
1 & 0 & 0 & 1\\
1 & 0 & 0 & -1\\
0 & 1 & 1 & 0\\
0 & -1 & 1 & 0
\end{pmatrix}.
\end{align}
We call it CNOT-H gate because its gate decomposition contains a CNOT gate and a Hadamard gate, as shown in Fig.~\ref{fig:decomposition}(b).

The matrix elements of $U_2$ fulfill the condition given in Eq.~(\ref{eq:condition_fix}), thus the GHZ state is a fixed point of the nonlinear transformation utilizing $U_2$. The success rate of the protocol for a GHZ input is 0.25.
However, some conditions in Eq.~(\ref{eq:condition_pure}) are violated by $U_2$, meaning that the protocol using $U_2$ cannot efficiently distill even the pure states.

The remarkable feature of the CNOT-H operator is the following. If the matrix elements of $M_\epsilon$ are all zeroes except for the elements $\langle 000|M_\epsilon|000\rangle$, $\langle 000|M_\epsilon|111\rangle$, $\langle 111|M_\epsilon|000\rangle$ and $\langle 111|M_\epsilon|111\rangle$, then
\begin{equation} \label{eq:PU2}
\mathcal{P}_{U_2}\left[\rho_\textrm{GHZ}+\epsilon M_\epsilon\right]=\rho_\textrm{GHZ} + \mathcal{O}(\epsilon^2).
\end{equation}
Thus, the protocol with $U_2$ is efficient in removing certain types of noise already in a single iteration. We note that this statement remains true, even if the two incoming noisy GHZ states are not perfect replicas of each other.

\section{Alternating double iteration protocol}
\label{ref_double_alternating}

As demonstrated in the previous section, a single iteration of the protocol cannot simultaneously correct coherent and incoherent errors in an efficient way. Therefore, in what follows, we will analyze the output after two consecutive iterations of the protocol. We will show that two iterations, using appropriately chosen unitary operators, can efficiently distill a GHZ state (i.e., eliminating the error in linear order) from any moderately distorted GHZ inputs.

Alternating schemes in distillation protocols, where the participants perform different local operations in subsequent iterations, have been studied previously. An early implementation of such a protocol is presented in Ref.~\cite{PhysRevA.57.R4075} for the distillation of white noise degraded GHZ states.

\subsection{Double iteration with $U_1$ and $U_2$}

\begin{figure}
\centering
\includegraphics[width=0.95\columnwidth]{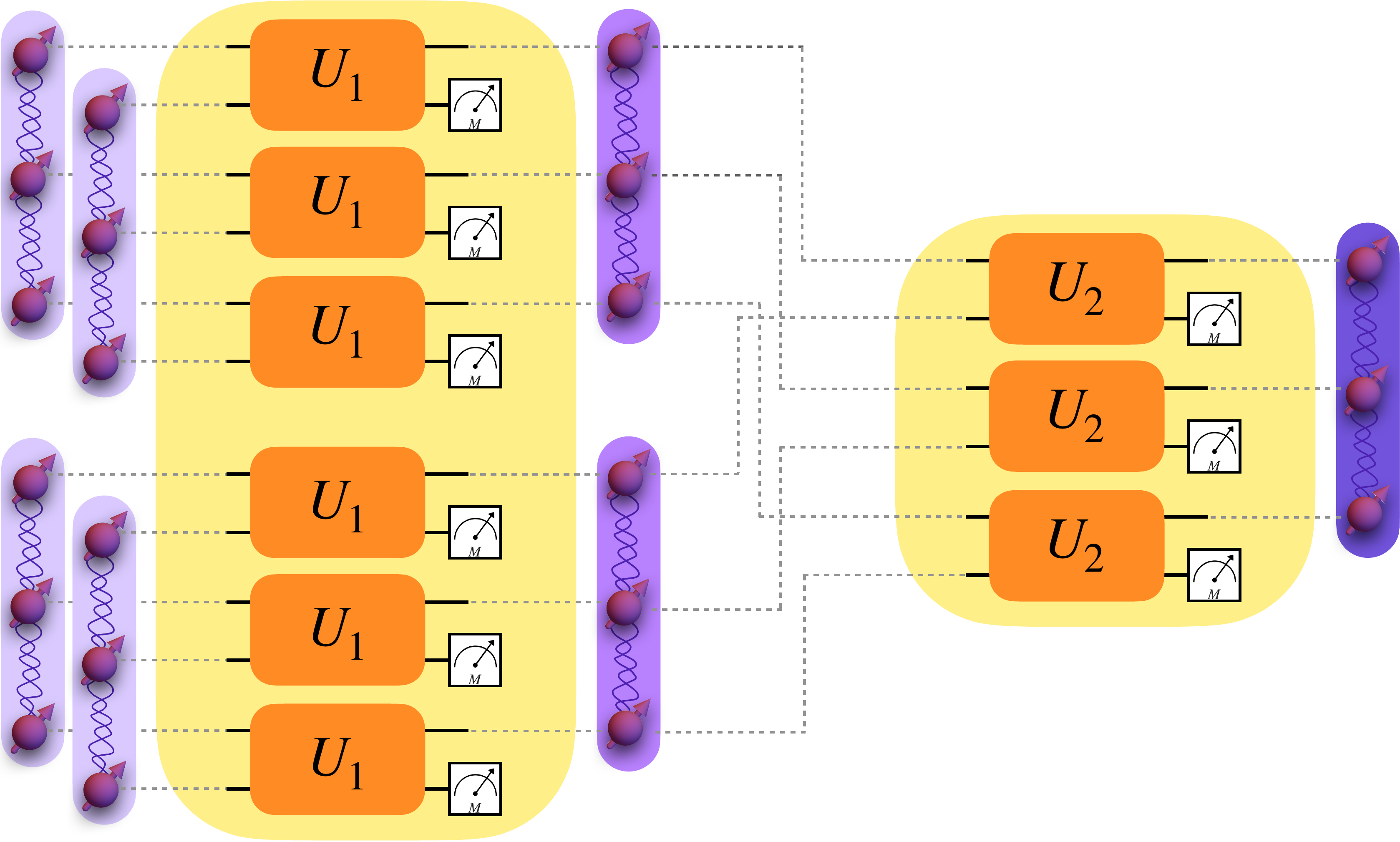}
\caption{\label{fig:alternating}Schematic diagram illustrating the alternating double iteration protocol employed for GHZ state distillation. In this scheme, the three parties perform different local operations ($U_1$ and $U_2$) in even and odd steps.}
\end{figure}

In this subsection -- as a main result of the paper -- we present the so-called alternating double iteration protocol, where the CNOT-X and CNOT-H operators are applied in the first and second iterations of the scheme, respectively, as shown in Fig.~\ref{fig:alternating}. The alternating double iteration protocol successfully eliminates the error in linear order:
\begin{equation} \label{eq:condition_alter}
\mathcal{P}_{U_2}\left[\mathcal{P}_{U_1}\left[\rho_\textrm{GHZ}+\epsilon M_\epsilon\right]\right]=\rho_\textrm{GHZ} +\mathcal{O}(\epsilon^2).
\end{equation}

The intermediate state of the alternating protocol, which is the output state of the protocol using $U_1$ and the input state of the protocol using $U_2$, is formulated by Eqs.~(\ref{eq:PU1}) and (\ref{eq:Mepsilon}). The form of the intermediate state satisfies the condition for Eq.~(\ref{eq:PU2}) to be valid, therefore, after the second application of the protocol (i.e., that with $U_2$), the GHZ state contains error only in quadratic order in $\epsilon$, as shown by Eq.~(\ref{eq:condition_alter}).
Note that further iterations of the alternating double iteration protocol always involve a consecutive application of $U_1$ and $U_2$.

For the rest of this section, by one iteration with the alternating protocol, we mean that we perform two iterations after each other, one using the CNOT-X operator and then one using the CNOT-H operator. A one-step iteration with the alternating protocol requires four input states, and the success rate is approximately $0.5\cdot 0.5\cdot 0.25=1/16$. For $n$ iterations of the alternating double iteration protocol, at least $4^n$ input states are needed, but the error of the output GHZ state is proportional to $\epsilon^{2^n}$. This means that the error of the GHZ state is reduced super-exponentially by increasing the number of iterations, which is a remarkable feature of the alternating protocol.

\begin{figure}
\centering
\includegraphics[width=0.95\columnwidth]{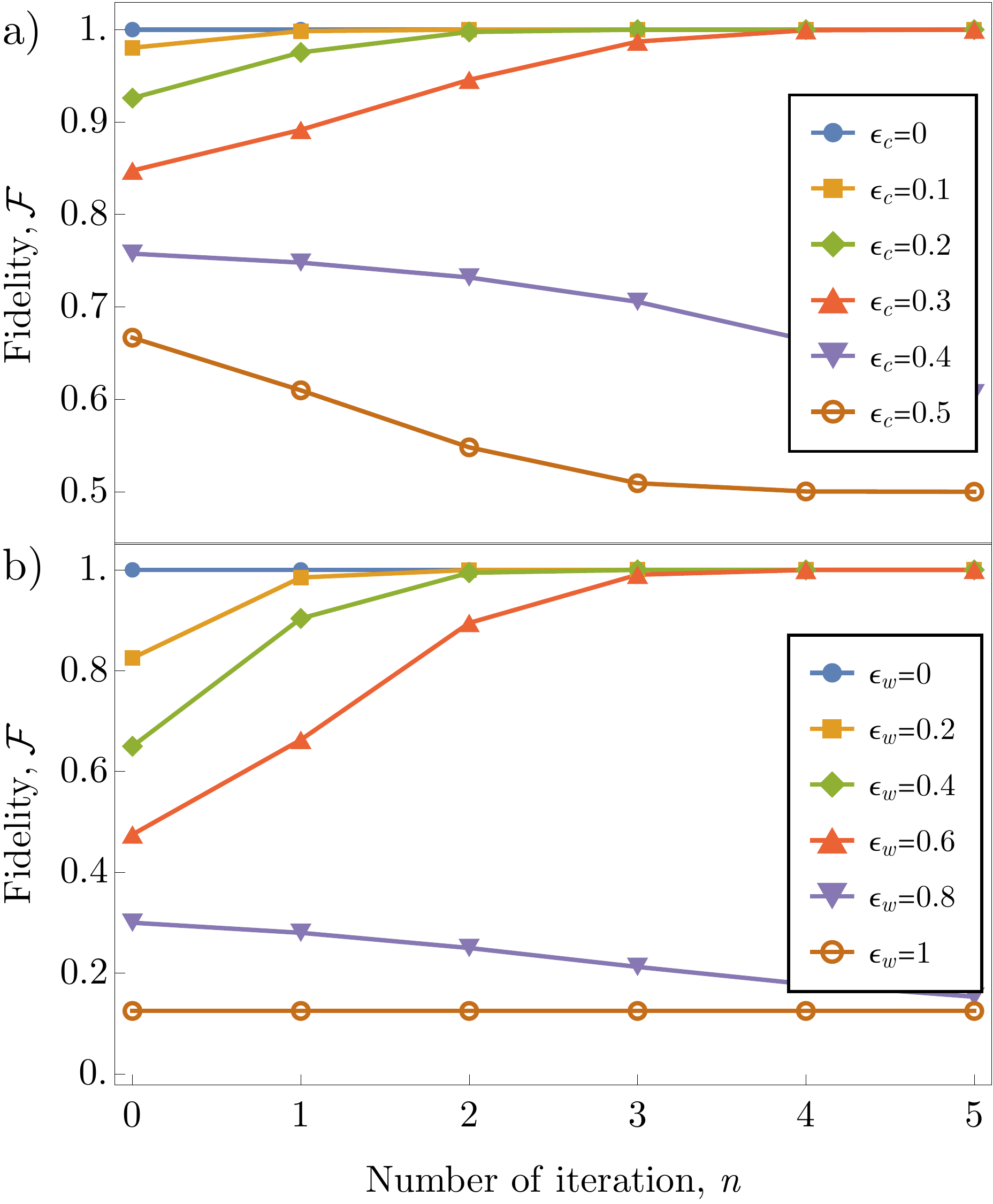}
\caption{\label{fig:noise}
The output fidelities obtained by the alternating double iteration protocol containing operators $U_1$ and $U_2$. The output fidelity is plotted as a function of the number of iterations (a) for pure-state inputs with coherent error represented as $\mathcal{N}\left(|GHZ\rangle+\epsilon_c |001\rangle + \epsilon_c |110\rangle\right)$, (b) for white noise affected GHZ states $\rho_\textrm{in} = (1-\epsilon_w)\rho_\textrm{GHZ}+\frac{\epsilon_w}{8} I$. In both cases, below a threshold value of the noise parameter $\epsilon_{c/w}$ the protocol successfully distills the GHZ state in a few iteration steps. Data points are connected by lines to aid visualization.
}
\end{figure}

\subsection{Numerical efficiency analysis}

In Fig.~\ref{fig:noise}, the efficiency of the alternating protocol is numerically investigated for two types of inputs: (a) a pure state and (b) a white noise-affected GHZ state, respectively. The efficiency is measured by the fidelity, which is the distance of the output state from the GHZ state:
\begin{equation}
\mathcal{F}=\langle \textrm{GHZ} | \rho_\textrm{out}|\textrm{GHZ}\rangle.
\end{equation}
In Fig.~\ref{fig:noise}(a), the fidelity is plotted as a function of the number of iterations for inputs, where the GHZ state is perturbed by extra terms of two of the computational basis states, e.g.,
\begin{equation}\label{eq:pure}
\mathcal{N}\left(|GHZ\rangle+\epsilon_c |001\rangle + \epsilon_c |110\rangle\right),
\end{equation}
where $\mathcal{N}$ is the normalization factor. In Fig.~\ref{fig:noise}(b), the protocol was applied on mixed states, where the GHZ state was perturbed by white noise
\begin{equation}\label{eq:mixed}
\rho_\textrm{in} = (1-\epsilon_w)\rho_\textrm{GHZ}+\frac{\epsilon_w}{8} I.
\end{equation}

The numerical examples imply the following main features: (i) For $\epsilon_{c/w}=0$ the fidelity remains $1$ because the GHZ state is a fixed point of the protocol, (ii) if the error is small ($\epsilon_{c/w}\ll 1$) the fidelity converges super-exponentially to $1$, (iii) if the error is below a threshold value being $\epsilon_c\cong 0.38$ in Fig.~\ref{fig:noise}(a) and $\epsilon_w\cong 0.75$ in Fig.~\ref{fig:noise}(b), the protocol can decrease the errors, and the fidelity converges to $1$. Above the threshold value, the fidelity does not converge to $1$ because the input state is outside of the attraction region of the fixed point of the dynamics, i.e., the GHZ state.

When coherent and incoherent errors are combined, the GHZ state remains an attractor; however, its basin of attraction decreases in size for both error types (Appendix~\ref{app:fates}, Fig.~\ref{fig:fatemap}). Outside this basin, the state converges to other fixed points of the protocol, such as a Bell pair with a separated third qubit or a mixture of $|000\rangle$ and $|111\rangle$. The specific fixed point reached depends on the nature of the error.

\subsection{Measurement error}

The distillation protocol involves measurements of the flag qubits. These measurements, however, may be erroneous. Measurement errors may occur in practice. For simplicity, these imperfections can be phenomenologically modeled such that, following a perfect projective measurement on the flag qubit, the opposite outcome is obtained with classical probability $p_m$. Subsequently, the parties share their possibly inaccurate measurement values, and building on this information, they decide to keep or discard the qubits.
The selective conditional measurement suppresses the noisy components of the input state. However, measurement errors introduce additional noise during the post-selection step. As a result, states that should potentially be discarded may be retained, and states that should be kept may be discarded. Consequently, it impairs the distillation ability of the protocol.
This appears in our calculation in a way that the result of the protocol is a probabilistic mixture of the outcome of the possible measurement scenarios.

\begin{figure*}
\centering
\includegraphics[width=\textwidth]{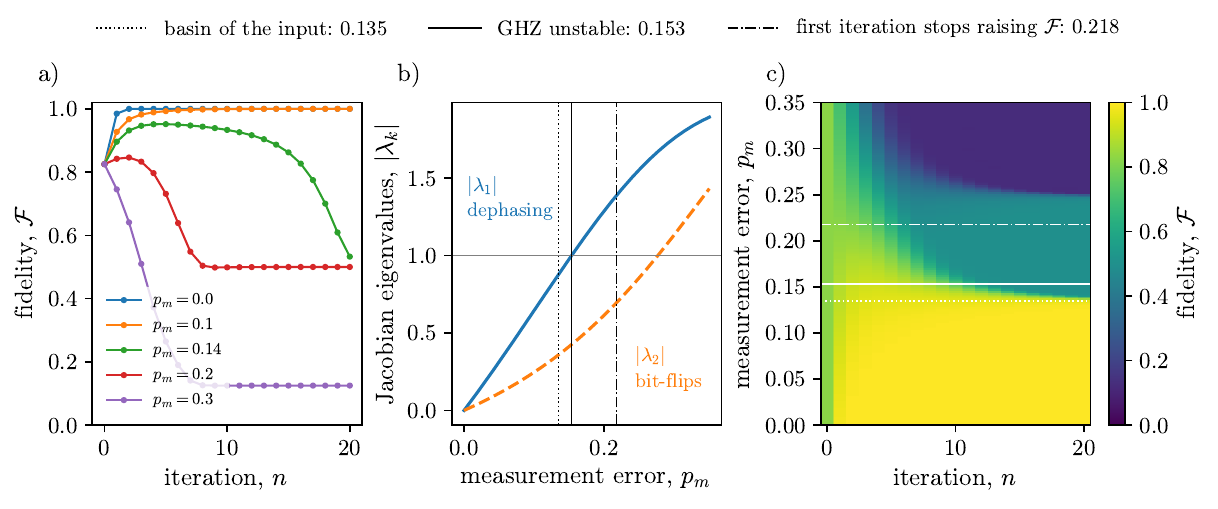}
\caption{\label{fig:measurementerror} The alternating double iteration protocol with measurement error $p_m$ for the input state $\rho_\textrm{in} = 0.8 \rho_\textrm{GHZ}+0.025 I$. (a) Output fidelity as a function of the number of iterations $n$ for various values of $p_m$. (b) Magnitude of the two largest eigenvalues of the Jacobian $J$ in Eq.~(\ref{eq:jacobian}), belonging to the dephasing ($|\lambda_1|$) and bit-flip errors ($|\lambda_2|$, threefold degenerate); the GHZ state is stable below the horizontal line $|\lambda|=1$. The vertical lines mark the threshold of the input ($p_m\cong 0.135$), the instability of the GHZ state ($p_m\cong 0.153$), and the value above which the first iteration no longer increases the fidelity ($p_m\cong 0.218$). (c) Output fidelity as a function of $n$ and $p_m$, with the thresholds of panel (b) as horizontal lines. 
}
\end{figure*}

Figures~\ref{fig:measurementerror}(a) and \ref{fig:measurementerror}(c) show the output fidelity for the input
$\rho_\textrm{in}=0.8\rho_\textrm{GHZ}+0.025I$. For $p_m<0.135$ the protocol reaches the GHZ state,
although more slowly than without measurement errors. For $0.135<p_m<0.218$ the behaviour is non-monotonic. The fidelity initially
increases during distillation but subsequently decreases, and the state ends in the mixture
$\frac12(|000\rangle\langle000|+|111\rangle\langle111|)$. For $p_m>0.218$ the fidelity decreases from
the first iteration, and for $p_m\gtrsim0.25$ the state converges to the maximally mixed state.

To explain this behaviour, we linearize one iteration $\mathcal{C}=\mathcal{P}_{U_2}\circ\mathcal{P}_{U_1}$
around the GHZ state, which remains a fixed point for any $p_m$,
\begin{equation}\label{eq:jacobian}
\mathcal{C}[\rho_\textrm{GHZ}+\epsilon M_\epsilon]=\rho_\textrm{GHZ}+\epsilon J[M_\epsilon]+\mathcal{O}(\epsilon^2).
\end{equation}
An error along an eigenvector of the Jacobian $J$ is multiplied by the eigenvalue $\lambda$ in every
iteration, so it shrinks for $|\lambda|<1$ and grows for $|\lambda|>1$. For $p_m=0$, $J=0$, which is the
elimination of noise to first order [Eq.~(\ref{eq:condition_alter})]. For $p_m>0$, the largest eigenvalue
[Fig.~\ref{fig:measurementerror}(b)] belongs to the dephasing error $|GHZ^-\rangle\langle GHZ^-|-\rho_\textrm{GHZ}$, while the
next one, belonging to the bit flip of any single qubit (threefold degenerate), stays below unity up to $p_m\cong0.28$.

Two types of thresholds must therefore be distinguished: the local stability threshold of the GHZ fixed point and the input-dependent thresholds characterizing the distillation dynamics. The local stability threshold is ($p_m \simeq 0.153$), where the largest eigenvalue of the Jacobian reaches unity. Above this value, even an arbitrarily small dephasing error grows under iteration.
For the particular input considered here, however, the protocol ceases to converge to the GHZ state already at ($p_m \simeq 0.135$), when the input leaves its basin of attraction. Nevertheless, the fidelity may still increase during the first iteration up to ($p_m \simeq 0.218$). In the intermediate range, ($0.135 \lesssim p_m \lesssim 0.218$), the fidelity initially improves as other error components decrease, but eventually drops as the dephasing component becomes dominant.

\subsection{Gate error}

The imperfections of the local unitary operations may be taken into account by the depolarizing error
model, where error channels are placed after every unitary operation. As a result, the two-qubit density matrix is transformed in the following way:
\begin{equation}
    \rho \rightarrow \left(1-\frac{16p_g}{15}\right)\rho + \frac{4p_g}{15} I_4.
\end{equation}
In this simple model, the imperfection of the local operations is characterized by a single parameter $p_g$.

\begin{figure}
\centering
\includegraphics[width=0.95\columnwidth]{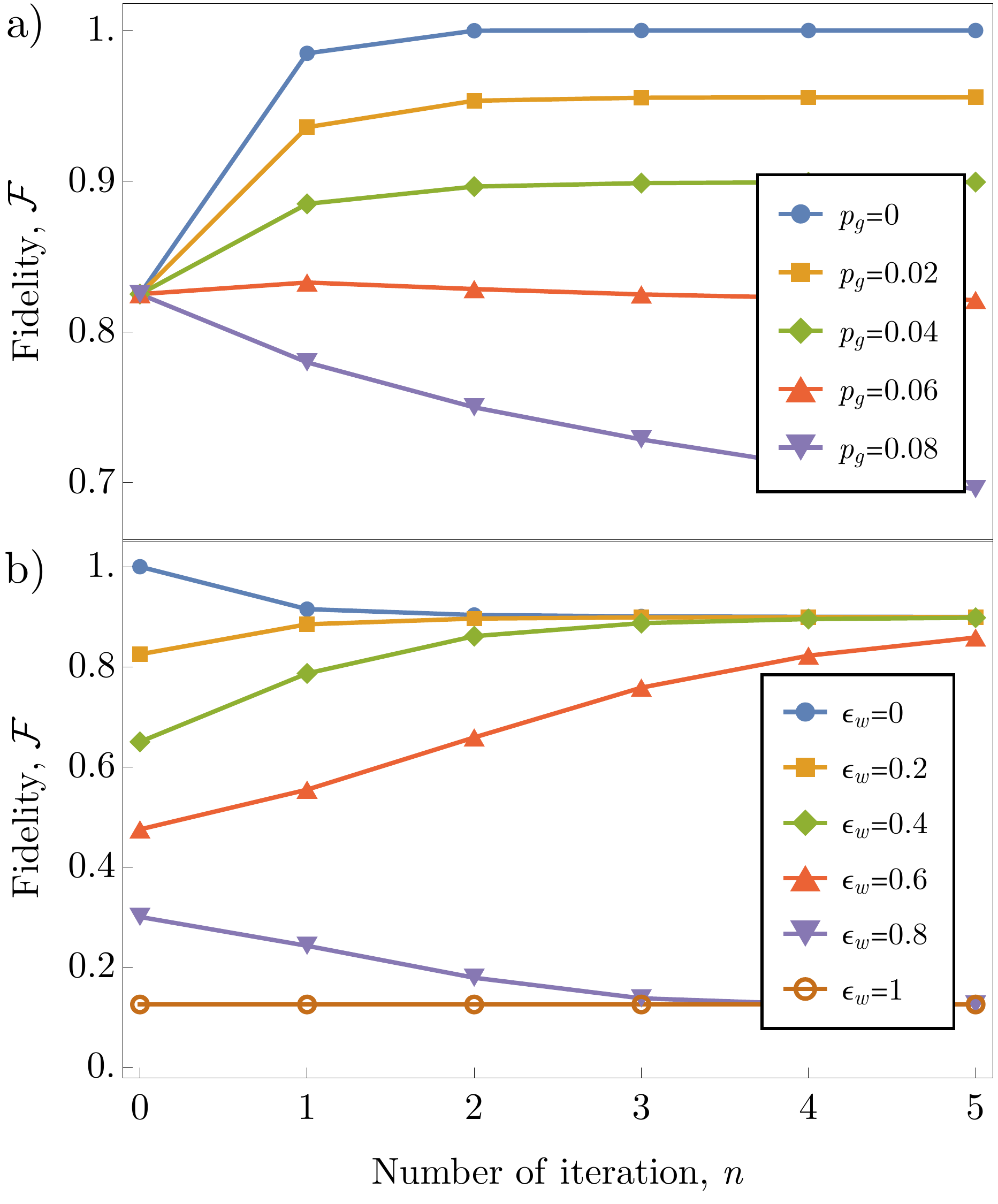}
\caption{\label{fig:gateerror} The output fidelity achieved by the alternating GHZ distillation protocol as a function of the iteration number (a) for different values of gate error $p_g$ when the input state is $\rho_\textrm{in} = 0.8 \rho_\textrm{GHZ}+0.025 I$, and (b) for various input states $\rho_\textrm{in} = (1-\epsilon_w) \rho_\textrm{GHZ}+\frac{\epsilon_w}{8} I$ at a fixed gate error of $p_g=0.04$. Data points are connected by lines to aid visualization.}
\end{figure}

In the case of non-perfect unitary operations, the GHZ state is no longer a fixed point of the non-linear transformation, hence perfect distillation cannot be achieved. However, at a small value of $p_g$ the fidelity is still enhanced by the protocol.
These statements are supported by Fig.~\ref{fig:gateerror}(a), where the output fidelity is plotted as a function of the
number of iterations for different values of gate error $p_g$. Our initial state is the same as in Fig.~\ref{fig:measurementerror}. For small values of $p_g$, which is $p_g<5\%$ in our example, the protocol improves the output fidelity. In Fig.~\ref{fig:gateerror}(b), we plot the output fidelity for the same inputs as in Fig.~\ref{fig:noise}(b), but with a fixed gate error of $p_g\cong 0.04$. These results imply that while the presence of gate error does not alter the size of the attraction basin around the GHZ state, but it reduces the convergence speed. Furthermore, a significant drawback is that the converged state is not a perfect GHZ state but is instead slightly distorted from it.
Our numerical results show that even a few percentage of gate errors significantly impair the distillation ability of the alternating protocol.

\section{Uniform double iteration protocol}
\label{ref_double_uniform}

As demonstrated in the previous section, utilizing two consecutive iterations of the protocol with different unitary operations results in an effective GHZ distillation scheme. However, it is not necessary to choose different operations during the successive iteration steps. In this section, we introduce the so-called uniform double iteration protocol, where the same unitary operator is applied in every iteration of the protocol.

\begin{figure}
\centering
\includegraphics[width=0.85\columnwidth]{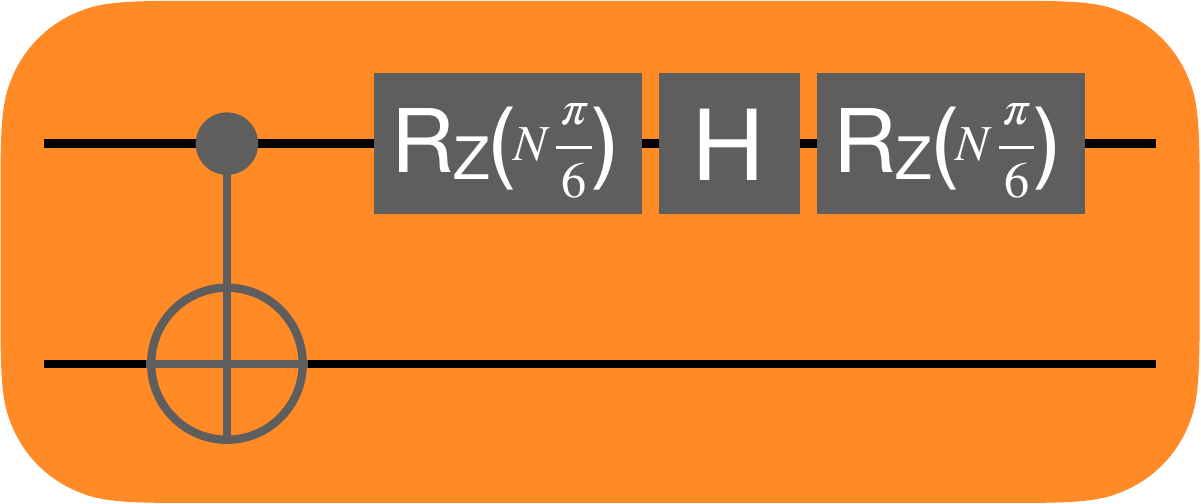}
\caption{\label{fig:uniform_decomposition} One possible gate decomposition of the unitary operator $U_3$ used in the uniform double iteration protocol. The bottom line represents the flag qubit.
 }
\end{figure}

Similarly as before, we impose the condition that the uniform double iteration protocol with a properly chosen unitary $U_3$ eliminates the noise in linear order
\begin{equation} \label{eq:condition_uniform}
\mathcal{P}_{U_3}\left[\mathcal{P}_{U_3}\left[\rho_\textrm{GHZ}+\epsilon M_\epsilon\right]\right]=\rho_\textrm{GHZ} +\mathcal{O}(\epsilon^2).
\end{equation}
An analytical approach to determine an appropriate operator $U_3$ leads to a system of nonlinear equations that is difficult to solve analytically. Therefore, in our previous work, we utilized a numerical method based on a variational quantum algorithm for this purpose. According to our numerical findings, we can construct a set of operators given by
\begin{align}\label{eq:U3}
U_3=\frac{1}{\sqrt{2}}
\begin{pmatrix}
e^{-iN\frac{\pi}{3}} & 0 & 0 & 1\\
0 & e^{-iN\frac{\pi}{3}} & 1& 0\\
1 & 0 & 0 & -e^{iN\frac{\pi}{3}}\\
0 & 1 & -e^{iN\frac{\pi}{3}} & 0
\end{pmatrix},
\end{align}
which satisfies the condition in Eq.~(\ref{eq:condition_uniform}) for any $N\in\mathbb{Z}$. For instance, a matrix we previously identified numerically (first line of Table 2 in Ref.~\cite{rozgonyi2024training}) coincides with Eq.~(\ref{eq:U3}) when $N=-2$. A possible gate decomposition of this operator $U_3$ utilizing CNOT, Hadamard, and RZ (rotation about the Z-axis) gates, is illustrated in Fig.~\ref{fig:uniform_decomposition}.

The GHZ state is a stable periodic point of the double iteration protocol, i.e., of the nonlinear transformation $\mathcal{P}_{U_3}\circ \mathcal{P}_{U_3}$, but not a fixed point of $\mathcal{P}_{U_3}$. We note that for odd-numbered iterations, the fidelity is given by
\begin{equation} \label{eq:fidU3odd}
\mathcal{F}=\langle \textrm{GHZ} | \mathcal{P}_{U_3}\left[\rho_\textrm{GHZ}\right]|\textrm{GHZ}\rangle=0.125,
\end{equation}
meaning that the state is significantly distant from the GHZ state. 
 The low fidelity in Eq.~(\ref{eq:fidU3odd}) does not reflect a loss of entanglement. For two GHZ inputs, the accepted
outcome of the first iteration occurs with probability $1/2$ and leaves the kept qubits in the state
\begin{equation}
  |\psi_N\rangle = V_N^{\otimes 3}\,|\mathrm{GHZ}\rangle ,
  \qquad
  V_N = e^{-iN\pi Z/6}\, H \, e^{-iN\pi Z/6},
\end{equation}
where $V_N$ is the single-qubit part of $U_3$ acting on the kept qubit after the CNOT gate (Fig.~\ref{fig:uniform_decomposition}).
Since this transformation consists of local unitary operations, the resulting state remains maximally entangled and is locally equivalent to the GHZ state written in the rotated local basis $\{V_N|0\rangle, V_N|1\rangle\}$, 
e.g.\ $|\psi_0\rangle = (|{+}{+}{+}\rangle + |{-}{-}{-}\rangle)/\sqrt{2}$ for $N=0$. 
The corresponding fidelity is
$\mathcal{F} = |\langle \mathrm{GHZ}|V_N^{\otimes 3}|\mathrm{GHZ}\rangle|^2 = [1+\sin^2(N\pi)]/8 = 1/8$.
The second iteration maps two copies of $|\psi_N\rangle$ back to $|\mathrm{GHZ}\rangle$ with probability $1/4$,
which yields the success probability $P = 1/16$ of a double iteration and makes the GHZ state a fixed
point of $\mathcal{P}_{U_3}\circ\mathcal{P}_{U_3}$ but not of $\mathcal{P}_{U_3}$.

\begin{figure}
\centering
\includegraphics[width=0.95\columnwidth]{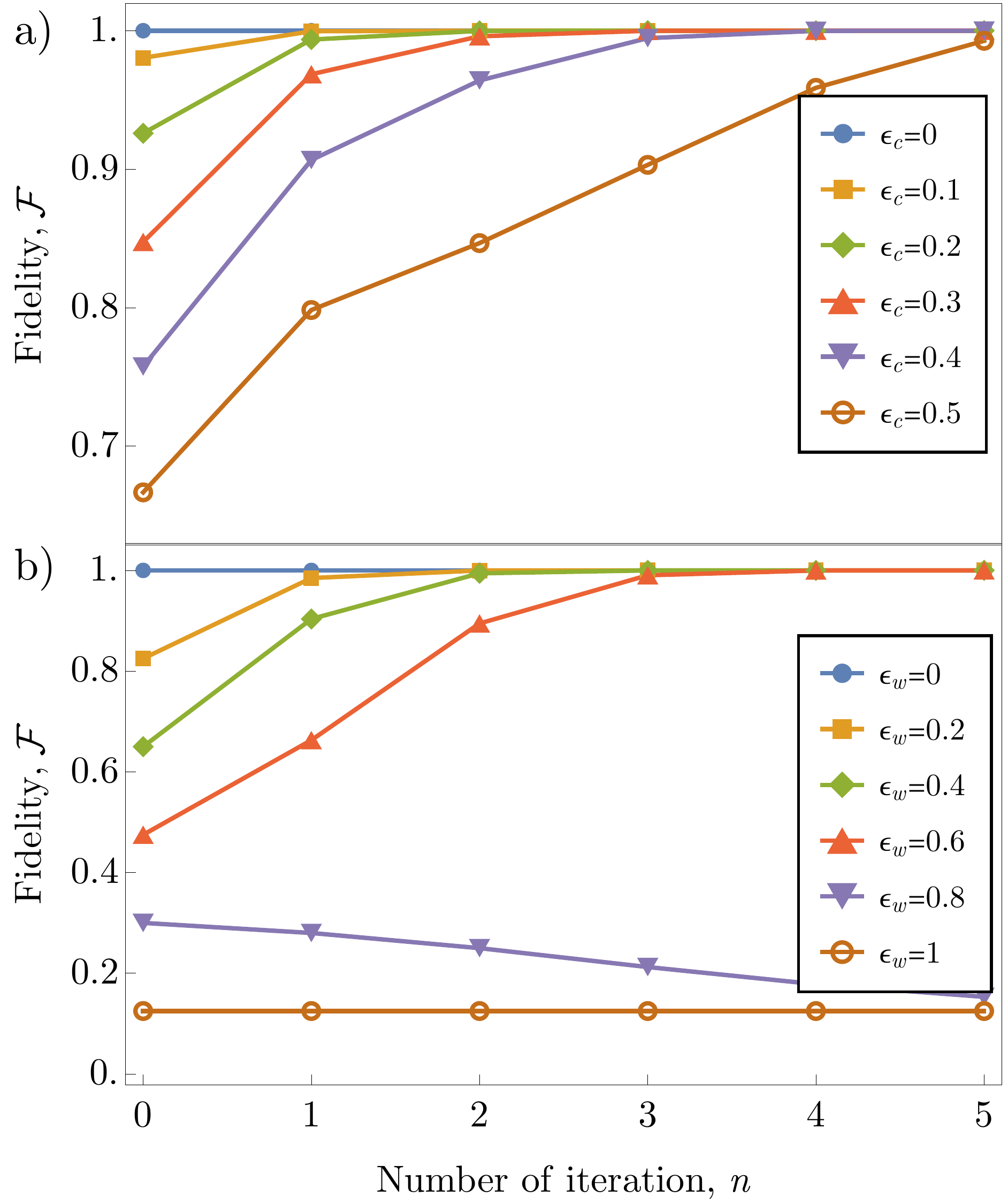}
\caption{\label{fig:uniform_noise}
The output fidelities obtained by the uniform double iteration protocol containing operators $U_3$. The output fidelity is plotted as a function of the number of double iterations (a) for pure-state inputs with coherent error represented as $\mathcal{N}\left(|GHZ\rangle+\epsilon_c |001\rangle + \epsilon_c |110\rangle\right)$, (b) for white noise affected GHZ states $\rho_\textrm{in} = (1-\epsilon_w)\rho_\textrm{GHZ}+\frac{\epsilon_w}{8} I$. In both cases, below a threshold value of the noise parameter $\epsilon_{c/w}$, the protocol successfully distills the GHZ state in a few iteration steps. Data points are connected by lines to aid visualization.
}
\end{figure}

In the next step, we analyze the tolerance of the uniform double iteration protocol utilizing $U_3$ with $N=-2$ in the case of input states slightly deviated from the GHZ states. We consider the same pure and mixed-state inputs as in the previous section [see Eqs.~(\ref{eq:pure}) and (\ref{eq:mixed})]. The output fidelities as a function of the number of double iterations are plotted in Fig.~\ref{fig:uniform_noise} for various values of the noise parameter $\epsilon_{c/w}$. One can see that the protocol effectively improves the fidelity for smaller values of $\epsilon_{c/w}$, and the fidelity converges superexponentially to its maximal value of $\mathcal{F}=1$. We note that choosing any other input states, containing either coherent or incoherent deviations from the GHZ state, would lead to similar results.

We can conclude that the uniform protocol using $U_3$ is equally efficient as the alternating protocol using $U_1$ and $U_2$ for distilling GHZ states perturbed by white noise. However, when applied to GHZ states affected by coherent noise, the uniform double iteration protocol is more effective than the alternating protocol, as demonstrated by the contrasting results in Fig.~\ref{fig:noise} and Fig.~\ref{fig:uniform_noise}.

We do not discuss in detail the distillation efficiency of the uniform protocol in the presence of measurement and gate errors, because we observe the same main features that we noted in the previous section for the alternating protocol. A precise discussion of the minor differences between the distillation capability of the alternating and uniform double iteration protocol could be a task for further work.

\section{Efficiency for coherent phase error}
\label{sec:practical}

In GHZ-state distribution, drifts of optical path lengths or interferometer arms, detuning, Zeeman, or AC-Stark shifts of the qubits, mismatched phase references of the nodes, or $ZZ$ crosstalk with spectator qubits can lead to a dominant coherent phase error
\begin{equation}\label{eq:phase}
|\Psi_\phi\rangle=\frac{1}{\sqrt{2}}\left(|000\rangle+e^{i\phi}|111\rangle\right),\quad \mathcal{F}_\textrm{in}=\cos^2(\phi/2).
\end{equation}
We examine the efficiency of our protocol for this case, compared to standard schemes such as that of Ref.~\cite{PhysRevA.57.R4075}.
In our scheme, only the measurement outcome $000$ is accepted, leading to a smaller success probability than in schemes similar to Ref.~\cite{PhysRevA.57.R4075}. For generic noisy inputs, such as GHZ states affected by white noise, this may lead to a higher resource consumption. For certain structured errors, such as the coherent phase error, however, our protocol has a significant advantage: the first iteration with $U_1$ can transform the input directly into the GHZ state, so the complete alternating sequence need not be executed. The smaller number of iterations (a single iteration in the present example) can compensate for the lower success rate. 

We assume no quantum memory: the $2^r$ copies needed for $r$ iterations are prepared together, and a failure in any iteration discards all of them. The expected number of raw copies per output state is then $C_r=2^r/Y_r$, where $Y_r=Y_{r-1}^2\,p_r$ ($Y_0=1$) is the probability that all iterations succeed and $p_r$ is the success rate of iteration $r$. Both protocols stop after the first iteration that reaches $\mathcal{F}\geq 0.999$.

\begin{figure}
\centering
\includegraphics[width=0.95\columnwidth]{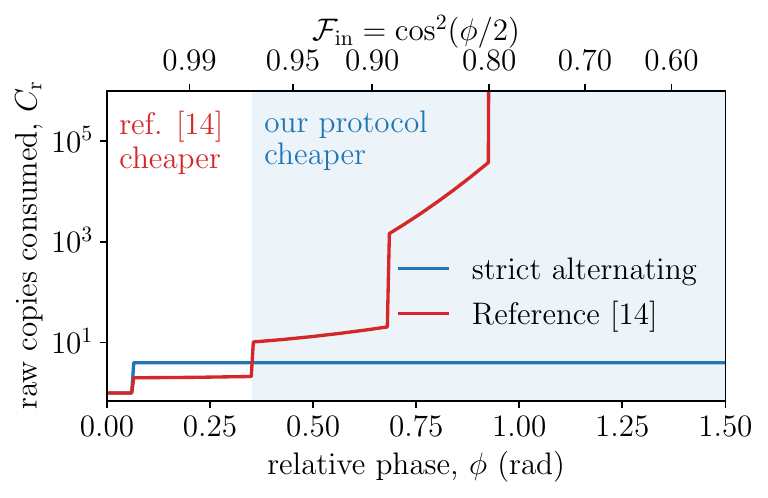}
\caption{\label{fig:practical} Expected number of raw copies consumed per output state with $\mathcal{F}\geq 0.999$ without quantum memory, for the input $|\Psi_\phi\rangle=1/\sqrt{2}\left(|000\rangle+e^{i\phi}|111\rangle\right)$.
Blue line represents our protocol with strict post-selection (outcome $000$ only), which here stops after the first iteration with $U_1$. Red line indicates the protocol of Ref.~\cite{PhysRevA.57.R4075}. Our protocol needs fewer copies in the shaded region, $\phi>0.35$.}
\end{figure}

Figure~\ref{fig:practical} and Table~\ref{tab:practical} show that our protocol needs 4 copies independently of $\phi$ (for $\mathcal{F}_\textrm{in}<0.999$), since the single iteration consumes two copies and succeeds with probability $1/2$, i.e., two attempts are needed on average. Since $|\Psi_\phi\rangle$ has the form of $|\Psi_0\rangle$ in Eq.~(\ref{eq:U1GHZ}), a single iteration with $U_1$ gives the exact GHZ state with success rate $1/2$: both accepted terms carry the factor $e^{i\phi}$ exactly once, so the phase becomes global. The protocol of Ref.~\cite{PhysRevA.57.R4075} accepts several outcomes and does not exploit the coherence of the error: its fidelities and success rates are the same as for the dephased input $\mathcal{F}_\textrm{in}\rho_\textrm{GHZ}+(1-\mathcal{F}_\textrm{in})|GHZ^-\rangle\langle GHZ^-|$. The error is therefore removed only gradually, and every additional iteration doubles the number of copies and lowers $Y_r$. This protocol is cheaper only for $\mathcal{F}_\textrm{in}>0.969$ ($\phi<0.35$), where a single iteration suffices, while for $\mathcal{F}_\textrm{in}<0.80$ it needs more than $10^{14}$ copies. If a small amount of white noise is present in addition to the phase error, $(1-\epsilon_w)|\Psi_\phi\rangle\langle\Psi_\phi|+\epsilon_w I/8$, a single iteration still suffices for $\epsilon_w\lesssim0.004$. For $\epsilon_w=0.01$--$0.03$, the complete double iteration is required, but our protocol remains more efficient for medium and large phases, $\phi\gtrsim0.8$ (last row of Table~\ref{tab:practical}).
We note that our tests found similar advantage for amplitude errors between $|000\rangle$ and $|111\rangle$.

\begin{table}
\caption{\label{tab:practical} Expected number of raw copies per output state with $\mathcal{F}\geq 0.999$ without quantum memory, for the input $(1-\epsilon_w)|\Psi_\phi\rangle\langle\Psi_\phi|+\epsilon_w I/8$ with the relative phase $\phi$ of Eq.~(\ref{eq:phase}) and white-noise admixture $\epsilon_w$, and input fidelity $\mathcal{F}_\textrm{in}$.}

\begin{ruledtabular}
\begin{tabular}{ccccc}
$\phi$ & $\epsilon_w$ & $\mathcal{F}_\textrm{in}$ & This work & Ref.~\cite{PhysRevA.57.R4075} \\
\hline
0.2 & 0 & 0.990 & 4 & 2.0 \\
0.5 & 0 & 0.939 & 4 & 13 \\
0.8 & 0 & 0.848 & 4 & $6.1\times 10^{3}$ \\
1.0 & 0 & 0.770 & 4 & $3.3\times 10^{16}$ \\
0.8 & 0.01 & 0.841 & 66 & $8.1\times 10^{3}$ \\
\end{tabular}
\end{ruledtabular}
\end{table}

\section{Conclusion}
\label{conclusion}

In this work, we have used analytical considerations to determine unitary operators for an iterated quantum protocol in an LOCC scheme, in order to effectively distill tripartite GHZ states. We have proved that a non-linear protocol, where the same two-qubit gate is applied in every iteration, cannot effectively improve the fidelity to the GHZ from step to step for slightly but arbitrarily distorted GHZ state inputs. However, we achieve a remarkable reduction of noise by applying double iteration protocols, when two consecutive iterational steps are performed. The unitary operators used in the double iteration protocol could be different as discussed in Sec.~\ref{ref_double_alternating} or the same as in Sec.~\ref{ref_double_uniform}. In both cases, the protocol efficiently distills the GHZ state from any coherently or incoherently distorted GHZ-state inputs, and the output fidelity to the GHZ state converges superexponentially to $\mathcal{F}=1$ as a function of the number of double iterations. From an experimental point of view, it can be an advantage that the unitary operations are relatively easy to gate decompose. Our distillation schemes tolerate around 10\% measurement error,
but the gate error impairs the maximally available fidelity, e.g., if the gate error is 2\%, the output fidelity converges to around 0.95.

Even though the investigated alternating and uniform double iteration protocols distill the GHZ state with similar efficiency, both have their own advantages. In the case of the alternating protocol, even the intermediate state is close to the GHZ state and the gate decomposition is simpler. For the uniform protocol, only one type of operation is repeated, and it tolerates larger coherent distortion. Deciding which protocol is worth choosing depends on the technical details of the platform used.

We envision that the generalization of our results will foster the design of distillation schemes for larger GHZ states that are especially important in distributed quantum computing or distributed error correction~\cite{de2024thresholds}.

\begin{acknowledgments}
This research was supported
by the Ministry of Culture and Innovation, and the National Research, Development and Innovation Office within the Quantum Information National Laboratory of Hungary (Grant No.~2022-2.1.1-NL-2022-00004), the KDP-2021 scheme (Grant No.~C1790232), by the J\'{a}nos Bolyai Research Scholarship of the Hungarian Academy of Science and by
the NKFIH through the OTKA Grants K~134437. T.K.\ and O.K.\ were supported by the NKFIH (TKP-2021-NVA-04). T.K.\ is grateful for the support from the ``Frontline'' Research Excellence Programme of the NKFIH (Grant No.~KKP133827). 
\'A.R.\ thanks the support from CeFEMA -- Centre of Physics and Engineering of Advanced Materials, through contracts UID/04540/2025, UID/PRR/04540/2025, and UID/PRR2/04540/2025 with the Portuguese Agência para a Investigação e Inovação AI2, and from LaPMET -- Laboratory of Physics for Materials and Emerging Technologies, through contract LA/P/0095/2020 with the Portuguese Agência para a Investigação e Inovação AI2.
\end{acknowledgments}

\section*{Data availability}
The code supporting the numerical and analytical findings of this research is available in our GitHub repository~\cite{Rozgonyi-GitHub}. No data was used for the research described in the article.


\appendix

\section{Combined errors and end states of the protocols}
\label{app:fates}

In this appendix, we consider inputs with coherent and incoherent errors together and identify the states that are reached outside the basin of attraction of the GHZ state.

\subsection{Combined coherent and incoherent errors}

\begin{figure*}
\centering
\includegraphics[width=0.9\textwidth]{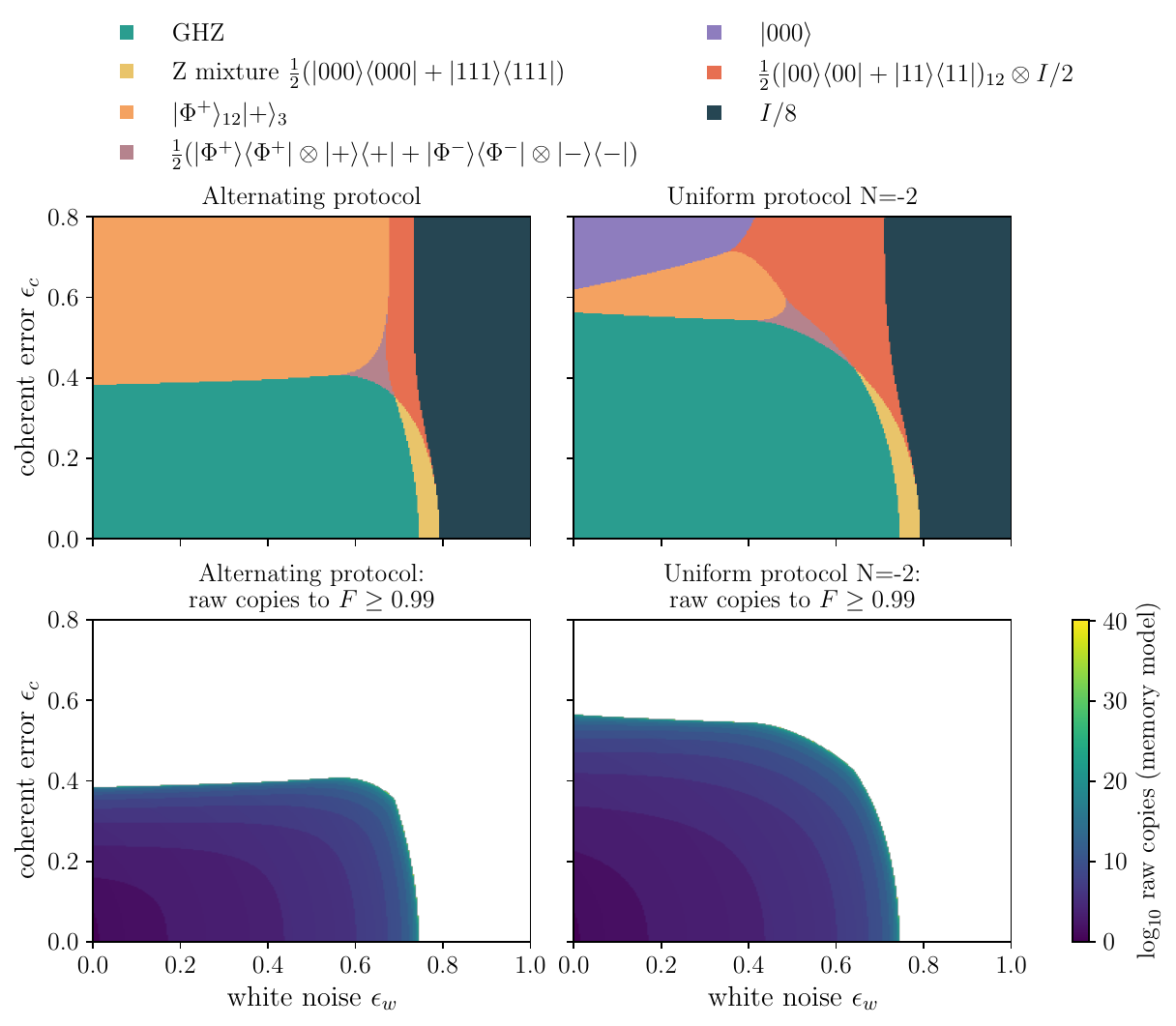}
\caption{\label{fig:fatemap} End states (upper panels) and resource cost (lower panels) for inputs with combined coherent and incoherent errors, Eq.~(\ref{eq:combined}), for the alternating protocol (left) and the uniform protocol with $N=-2$ (right). The uniform protocol reaches $|\Phi^+\rangle_{12}|+\rangle_3$ and its mixed counterpart in a locally rotated frame. Lower panels: expected number of raw copies needed to reach $\mathcal{F}\geq0.99$ within the GHZ basin; unlike in Sec.~\ref{sec:practical}, a quantum memory is assumed here, i.e., successful outputs are stored and paired again, so that a failed iteration discards only its own two input states.}
\end{figure*}

We combine the errors of Eqs.~(\ref{eq:pure}) and (\ref{eq:mixed}),
\begin{equation}\label{eq:combined}
\rho_\textrm{in}=(1-\epsilon_w)|\Psi_c\rangle\langle\Psi_c|+\frac{\epsilon_w}{8}I,\qquad
\mathcal{F}_\textrm{in}=\frac{1-\epsilon_w}{1+2\epsilon_c^2}+\frac{\epsilon_w}{8},
\end{equation}
where $|\Psi_c\rangle=\mathcal{N}\left(|GHZ\rangle+\epsilon_c|001\rangle+\epsilon_c|110\rangle\right)$. 
Figure~\ref{fig:fatemap} shows the result for the alternating protocol and for the uniform protocol with $N=-2$ in Eq.~(\ref{eq:U3}).

For the alternating protocol, the GHZ basin is bounded by $\epsilon_c\cong0.384$ for $\epsilon_w=0$ and by $\epsilon_w\cong0.745$ for $\epsilon_c=0$. The bound in $\epsilon_w$ decreases only slowly with $\epsilon_c$, 
i.e., the two types of error interact only weakly over most of the basin. The uniform protocol has the same bound in $\epsilon_w$ but tolerates coherent errors up to $\epsilon_c\cong 0.564$. Outside the basin, the state converges to one of the following fixed points, where $|\Phi^\pm\rangle=(|00\rangle\pm|11\rangle)/\sqrt{2}$:
\begin{enumerate}
\item Bell pair and separated qubit: $|\Phi^+\rangle_{12}|+\rangle_3$, with $\mathcal{F}=1/2$.
\item Its mixed counterpart: $\frac{1}{2}\left(|\Phi^+\rangle\langle\Phi^+|\otimes|+\rangle\langle+|+|\Phi^-\rangle\langle\Phi^-|\otimes|-\rangle\langle-|\right)$, with $\mathcal{F}=1/2$.
\item Mixture of $|000\rangle$ and $|111\rangle$: $\frac{1}{2}\left(|000\rangle\langle000|+|111\rangle\langle111|\right)$, with $\mathcal{F}=1/2$.
\item Classical pair and maximally mixed qubit: $\frac{1}{2}\left(|00\rangle\langle00|+|11\rangle\langle11|\right)_{12}\otimes I/2$, with $\mathcal{F}=1/4$.
\item Maximally mixed state: $I/8$, with $\mathcal{F}=1/8$.
\item Product state: $|000\rangle$, with $\mathcal{F}=1/2$, reached only by the uniform protocol.
\end{enumerate}
The uniform protocol reaches states 1 and 2 in a locally rotated frame.

The coherent error in Eq.~(\ref{eq:pure}) can be written as $\epsilon_c(|001\rangle+|110\rangle)=\sqrt{2}\,\epsilon_c X_3|GHZ\rangle$, where $X_3$ is the Pauli operator acting on the third qubit. Consequently, the input state contains a component directed towards another attractor, $|\Phi^+\rangle_{12}|+\rangle_3=(|GHZ\rangle+X_3|GHZ\rangle)/\sqrt{2}$. The coherence between these components produces a nonzero polarization of the third qubit, $\langle X_3\rangle=2\sqrt{2}\,\epsilon_c(1-\epsilon_w)/(1+2\epsilon_c^2)$. Above the coherent-error threshold, the protocol amplifies this polarization, which eventually separates the third qubit.
Interestingly, adding white noise can suppress this effect and restore convergence to the GHZ state, although it lowers the input fidelity, it changes the position of the initial state relative to the basins of attraction. For example, for the alternating protocol at $\epsilon_c=0.40$, the GHZ state is reached for $0.443<\epsilon_w<0.617$.

\squeezetable  
\begin{table}[t]
\caption{\label{tab:fates} End states reached for coherent errors along the direction $|d\rangle$, which replaces $|001\rangle+|110\rangle$ in Eq.~(\ref{eq:combined}), for $0\leq\epsilon_c\leq0.8$ and $0\leq\epsilon_w\leq1$, by the alternating protocol (A) and the uniform protocol with $N=-2$ (U). Columns: $|d\rangle=|001\rangle+|110\rangle$, as in Eq.~(\ref{eq:pure}) (its permutations and $i(|001\rangle+|110\rangle)$ give the same end states); $|001\rangle$; amplitude (Ampl.) error $|000\rangle-|111\rangle$; phase error $i(|000\rangle-|111\rangle)$; 40 random directions. End states are identified up to local unitaries and permutations of the qubits, using the spectra of the state and of its reduced states; e.g., ``pure product'' contains $|000\rangle$ and $|{+}{+}{+}\rangle$, ``Bell pair $\otimes$ qubit'' contains $|\Phi^+\rangle_{12}|+\rangle_3$, ``mixed'' denotes its mixed counterpart, and ``classical pair'' is $\frac{1}{2}(|00\rangle\langle00|+|11\rangle\langle11|)$.}

\begin{ruledtabular}
\begin{tabular}{lccccc}
End state & Eq.~(\ref{eq:pure}) & $|001\rangle$ & Ampl. & Phase & Random \\
\hline
GHZ & A,U & A,U & A,U & A,U & A,U \\
Mixture of $|000\rangle$, $|111\rangle$ & A,U & A,U & A,U & A,U & A,U \\
Bell pair $\otimes$ qubit & A,U & U & & & A,U \\
Bell pair $\otimes$ qubit, mixed & A,U & A,U & & & A,U \\
Classical pair $\otimes$ $I/2$ & A,U & A,U & & & A,U \\
$I/8$ & A,U & A,U & A,U & A,U & A,U \\
Pure product & U & U & U & U & A,U \\
Other & & & & & A,U \\
\end{tabular}
\end{ruledtabular}
\end{table}

\subsection{General properties of the end states}

The end states follow from four exact properties of a single iteration $\mathcal{P}_U$:
(i) all parties apply the same unitary and the accepted outcome $000$ is symmetric, so $\mathcal{P}_U[\Pi\rho\Pi^\dagger]=\Pi\,\mathcal{P}_U[\rho]\,\Pi^\dagger$ for any permutation $\Pi$ of the qubits;
(ii) for a product input $\rho_A\otimes\rho_B$, where $A$ and $B$ are complementary sets of parties, the two copies, the local unitaries, and the projection of the flag qubits factorize, so $\mathcal{P}_U[\rho_A\otimes\rho_B]=\mathcal{P}_U^{(A)}[\rho_A]\otimes\mathcal{P}_U^{(B)}[\rho_B]$, where $\mathcal{P}_U^{(A)}$ denotes the same protocol acting on the parties in $A$ only;
(iii) $I/8$ is a fixed point, because $U^{\otimes3}$ leaves the maximally mixed state of the two copies invariant;
(iv) pure inputs remain pure (Sec.~\ref{sec_setup}).

Consequently, the possible end states are fixed points of the protocol and do not depend on the input; the input only determines which of them is reached. By (i), permuting the error permutes the end state; since Eq.~(\ref{eq:pure}) is symmetric under the exchange of qubits 1 and 2, only qubit 3 can separate. By (ii), products of fixed points of the protocol acting on fewer parties are fixed points as well; for instance, $|\Phi^+\rangle_{12}|+\rangle_3$ is a fixed point because $|\Phi^+\rangle$ and $|+\rangle$ are fixed points of the alternating protocol acting on two parties and on a single party, respectively. By (iv), mixed end states require incoherent noise.

\subsection{Other coherent errors}

To test how the end states depend on the type of error, we replaced $|001\rangle+|110\rangle$ in $|\Psi_c\rangle$ by other directions $|d\rangle$ of the same norm, orthogonal to $|GHZ\rangle$, including 40 random ones, and scanned the range of Fig.~\ref{fig:fatemap}.
Table~\ref{tab:fates} lists the end states reached. The permuted errors $|010\rangle+|101\rangle$ and $|100\rangle+|011\rangle$ give the same end state as Eq.~(\ref{eq:pure}) at every point, with qubit 2 or 1 separated instead of qubit 3, as required by property (i). Errors within the span of $|000\rangle$ and $|111\rangle$ never lead to a Bell pair with a separated qubit; for $\epsilon_w=0$, the alternating protocol removes a relative phase exactly, cf.~Eq.~(\ref{eq:U1GHZ}). Random errors reach all of the above end states, with any of the three qubits separated, as well as a few further states, e.g., $|{+}{+}{+}\rangle$.

\bibliographystyle{apsrev4-2}
\nocite{apsrev42Control}  
\bibliography{paper}

\end{document}